\documentclass[journal,fleqn]{IEEEtran}
\pdfoutput=1

\usepackage[utf8]{inputenc}
\usepackage[T1]{fontenc}

\usepackage{authblk}

\usepackage{graphicx}
\usepackage{framed}
\usepackage{float}
\usepackage{balance}
\usepackage[inline]{enumitem}

\usepackage{listings}
\usepackage{xcolor}

\definecolor{codegreen}{rgb}{0,0.6,0}
\definecolor{codegray}{rgb}{0.5,0.5,0.5}
\definecolor{codepurple}{rgb}{0.58,0,0.82}

\lstdefinestyle{mystyle}{
    commentstyle=\color{codegreen},
    keywordstyle=\color{magenta},
    numberstyle=\tiny\color{codegray},
    stringstyle=\color{codepurple},
    basicstyle=\ttfamily\footnotesize,
    breakatwhitespace=false,         
    breaklines=true,                 
    captionpos=b,                    
    keepspaces=true,                 
    numbers=left,                    
    numbersep=5pt,                  
    showspaces=false,                
    showstringspaces=false,
    showtabs=false,                  
    tabsize=2
}

\lstdefinelanguage{XML}
{
  basicstyle=\ttfamily\footnotesize,
  morestring=[b]",
  morecomment=[s]{<?}{?>},
  stringstyle=\color{codegreen},
  identifierstyle=\color{blue},
  keywordstyle=\color{red},
  morekeywords={ID, resource}
}

\usepackage[font=normalsize]{caption}
\usepackage{algorithm}
\usepackage{algorithmic}

\usepackage{mathtools}

\usepackage[sorting=none]{biblatex}
\author[ ]{Duncan Deveaux$^{\dagger}\negmedspace$}
\author[ ]{Takamasa Higuchi$^{\ddagger}\negmedspace$}
\author[ ]{Seyhan Uçar$^{\ddagger}\negmedspace$}
\author[ ]{Jérôme Härri$^{\dagger}\negmedspace$}
\author[ ]{Onur Altintas$^{\ddagger}\negmedspace$}

\affil[ ]{$^{\dagger}$EURECOM, Campus SophiaTech, 450 route des Chappes, 06904 Sophia-Antipolis, France}
\affil[ ]{E-mail: \texttt {\{deveaux,haerri\}@eurecom.fr}}

\affil[ ]{$^{\ddagger}$InfoTech Labs, Toyota Motor North America R\&D, Mountain View CA, USA}
\affil[ ]{E-mail: \texttt {\{takamasa.higuchi,seyhan.ucar,onur.altintas\}@toyota.com}}

\title{A Definition and Framework for Vehicular Knowledge Networking}

\begin{document}


\maketitle

\begin{abstract}
To operate intelligent vehicular applications such as automated driving, machine learning, artificial intelligence and other mechanisms are used to abstract from information what is commonly referred to as knowledge. Defined as a state of understanding obtained through experience and analysis of collected information, knowledge is promising for vehicular applications. However, it lacks a unified framework to be cooperatively created and shared to achieve its full potential. This paper investigates on the meaning and scope of knowledge applied to vehicular networks, and suggests a structure for vehicular knowledge description, storage and sharing. Through the example of passenger comfort-based rerouting, it exposes the potential benefits for network load and delay of such knowledge structuring.

\end{abstract}

\IEEEpeerreviewmaketitle

\section{Introduction}

\IEEEPARstart{O}{ver} the last decade, we have witnessed the evolution of vehicular networking from 'Vehicular Ad-Hoc Networks' (VANETs), enabling spontaneous direct communications between vehicles, to 'Connected Vehicles' generalizing information exchange among vehicles and infrastructure. Vehicular networking has been developed as an enabler of innovative applications intended to improve traffic safety, reduce congestion and even provide infotainment on-board. Early applications were designed to only provide information to drivers, delegating any decision making to them.
However, in recent ambitious applications, such as automated driving or platooning, simple information treatment and forwarding mechanisms are not sufficient anymore. Instead, decision-making is based on models of the environment built and improved from increasingly large and diverse sets of input information. Models are designed to learn from experience rather than react to static input signals. In this context, the current standards of vehicular networking, designed for the exchange of static information alone, are no longer optimal. As the individual piece of information becomes meaningless compared with the model that was extracted from it, the scope of vehicular networks is gradually shifting to the creation and exchange of \emph{knowledge} models, as opposed to the sum of their training input.

\begin{figure}
    \centering
    \includegraphics[width=\columnwidth]{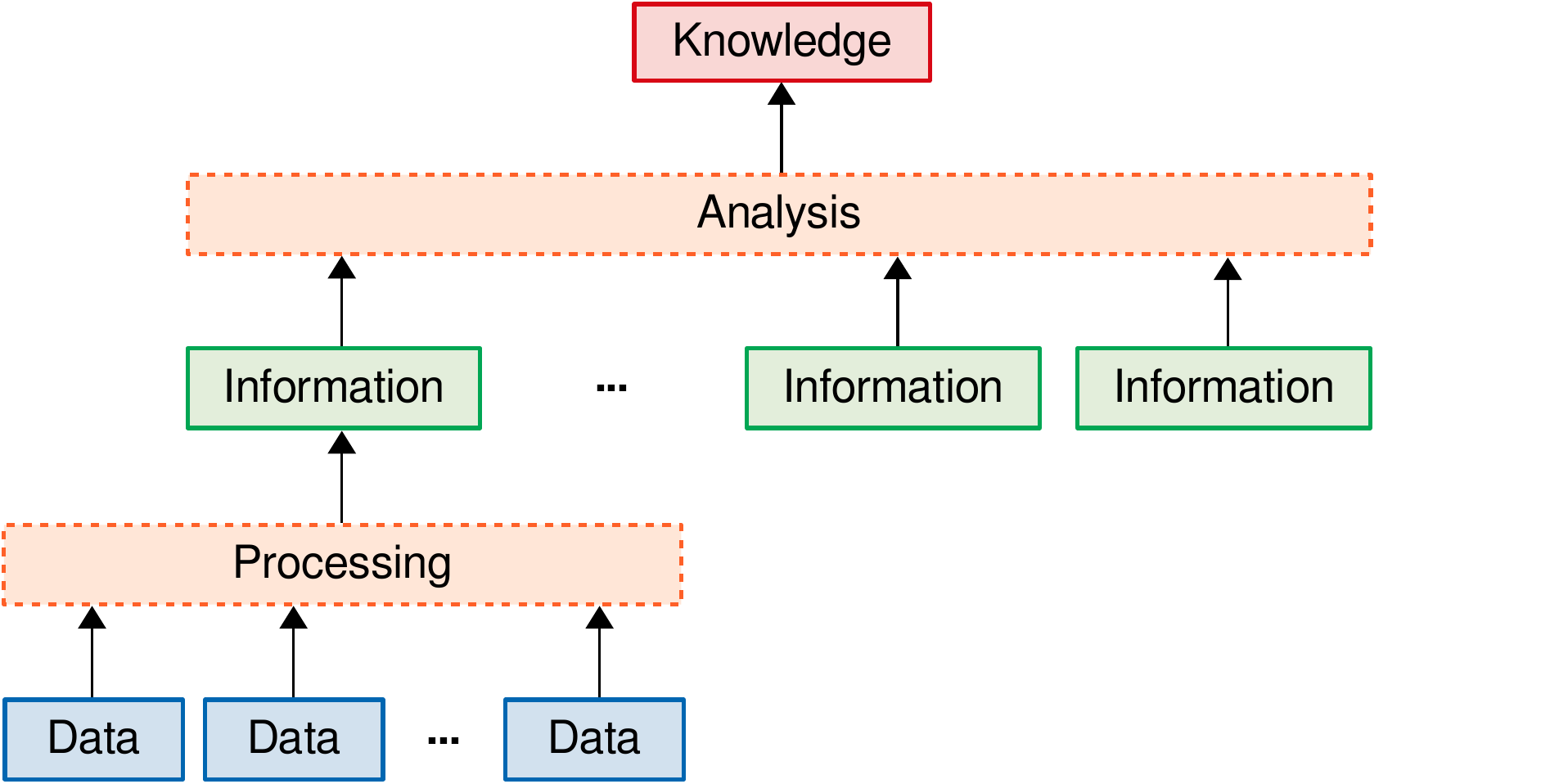}
    \caption{The Relationship between Data, Information and Knowledge}
    \label{fig:data_info_knowledge}
\end{figure}

The existing literature in the information science domain covers conceptual definition of data, information and knowledge \cite{zins2007}. In this paper, for the sake of clarity, we make similar distinctions among these categories. As shown in Figure~\ref{fig:data_info_knowledge}, the most fundamental element is data, that we define as an atomic value with a unit e.g. (5km). Next, information is built by aggregating pieces of data that describe a situation, e.g. (15:00, 30kph) a vehicle's speed at a given time. On top of information lies knowledge, which describes general patterns and relationships obtained through the analysis of sets of information. For example, clustering or classification algorithms can be used to extract hidden relationships within a set. Considering vehicles' pollutant emissions and speed through a day, a piece of extracted knowledge could be an estimate of the extra emissions linked to peak time.

Various techniques, such as Artificial Intelligence (AI), Machine Learning (ML) or Formal Language (FL) have been used to extract knowledge in vehicular contexts through the analysis of various sources of information~\cite{yeLiang2018}. For example, Ruta~{\em et al.}~\cite{ruta2018} composed {\em knowledge} to recognize a high level context of driving by using sensor information from multiple cars in a common geographical area. Qi, Wang~{\em et al.}~\cite{qi2019} applied both learning algorithms and edge computing units offloading to provide optimal caching of high level connected driving services to vehicles, including image auto annotation or locally relevant recommendations yielding. Khan~{\em et al.}~\cite{khan2019} applied deep learning to learn the transmission patterns of neighboring vehicles and paved the way for fewer packet collisions.

Regardless of the technique, extracting {\em knowledge} from information is a complex and expensive process, and the generated {\em knowledge} may be beneficial to other vehicles. So far each vehicle remains autonomous for its {\em knowledge} building, which requires highly specialized algorithms and a large amount of input information, potentially sourced from multiple different vehicles. This can be seen as a significant overhead considering that {\em knowledge} can be shared and not individually recreated. As a reaction, research has recently been focused on defining a knowledge-centric approach to networking, where information would not be the main focus anymore. Instead, knowledge would be created by nodes in the network, and directly stored and shared among them.
Wu~{\em et al.}~\cite{wu2019} described the concept of a knowledge-centric networking framework, separated into the three building blocks of knowledge creation, composition and distribution. A literature survey on means of creating and distributing knowledge is performed. However, the concept of knowledge remains abstract and its implementation or format is left for future work.

The structure and understanding of {\em knowledge} needs to be harmonized. Knowledge must move, follow and adapt in symbiosis with information as well as other pieces of knowledge. 
Finally, existing information networking mechanisms such as Information-Centric Networking (ICN) lack semantic annotations describing such context, which prevails them to uniquely locate and acquire the required knowledge~\cite{yao2019}. The naming, storage and dissemination of {\em knowledge} thus must integrate semantic annotations, which in turn need to be harmonized.

In this paper we aim to present Vehicular Knowledge Networking (VKN), a knowledge-centric networking framework applied to vehicular networks. We suggest a common architecture for knowledge description, needed for subsequent storage, composition and exchange with other connected vehicles. VKN is a framework that makes performance improvements in other applications possible. In a conceptual passenger comfort-based rerouting application, we show a potential load and delay reduction in vehicular networks through cooperative knowledge building and sharing.

The rest of this article is organized as follows: Section~\ref{sec:infovsknowledge} introduces information treatment standards and defines the scope of \emph{knowledge} in vehicular networks. Based on this understanding, Section~\ref{sec:vkn} describes a structure for knowledge description, storage and distribution. In Section~\ref{sec:application}, an application of the concept shows potential load and delay improvement for the network. Section~\ref{sec:opportunities} finally points out the potential research applicability behind VKN, while Section~\ref{sec:conclusion} summarizes the article.

\section{Vehicular Information and Vehicular Knowledge}
\label{sec:infovsknowledge}

In this section, we first describe the current forms of information in vehicular networks, as well as various standards for information storing and sharing. Then, we build on this understanding to define a format for vehicular knowledge representation.

\subsection{Information in Vehicular Networks}

Nodes of the vehicular network may exchange diverse types of information, including but not limited to:
\begin{itemize}
\item Safety notifications, e.g. accidents or road condition.
\item Vehicle state information and sensor measurements.
\item Navigation information, e.g. maps, road or parking data.
\item Information on topics such as weather or traffic flow.
\item Road-related information, e.g. gas station opening times.
\item Multimedia contents for user infotainment.
\end{itemize}

Various standards to represent and store information on-board vehicles have been developed in the past years.
In ETSI standards, the storage of content in connected vehicles is performed inside the \emph{Local Dynamic Map (LDM)} information base~\cite{ETSIstandard_LDM}, itself divided into four layers:
\begin{enumerate}
    \item Permanent static data, i.e. map data.
    \item Temporary static data, i.e. roadside infrastructure.
    \item Temporary dynamic data, e.g. roadblock, signal phase.
    \item Highly Dynamic data, e.g. vehicles, pedestrians.
\end{enumerate}


Although the LDM provides a standard approach for storing information, it does not provide any standard for information naming. Generally, any information may be stored in the LDM as long as it is labeled with a space-time area of relevance. This can lead to a lack of interoperability between the contents generated by different providers. To tackle this issue, semantic standards have been developed in order to provide nodes with a "common language" and avoid redundancy. For example, Syzdykbayev {\em et al.}~\cite{syzdykbayev2019} defined an ontology for interactions between the different actors of vehicular networks and the infrastructure. Moreover, the Vehicle Signal Specification (VSS), later extended to the VSSo ontology~\cite{vsso2018}, provides a standard way to address the state of the inner components of vehicles, e.g. steering wheels or windows opening.

Finally, after it has been created and stored, information is spread within the vehicular network. In most vehicular applications, nodes care more about the information itself than about the host it originated from. Routing algorithms have thus been developed that focus on the information being shared rather than on its actual host and location in the network. Information-Centric Networking (ICN) is a networking paradigm that may be suitable for some vehicular applications. Rather than sending a request to a specific host, using ICN, a vehicle disseminates a request to fetch a specific information identified by a unique content name.

\subsection{Vehicular Knowledge Representation}

In next generation vehicular networks, we envision a shift away from the information-based architecture to the benefit of knowledge. As with information, mechanisms must be developed to create, compose, store and distribute knowledge, as introduced in the survey by Wu {\em et al.}~\cite{wu2019}. In order to go further and based on definitions of knowledge, we propose a formal structure for knowledge representation. In turn, it paves the way for the development of knowledge storage and distribution mechanisms.

We understand knowledge as a piece of abstract content obtained from the analysis of a larger set of information~\cite{zins2007}.
Knowledge can be extracted from information using machine learning algorithms, divided into three classes. Supervised learning is a technique applied to classification or regression. A model is trained based on a number of samples of the form: (information, class) for classification -- or (information, value) for regression. This step, called model training, allows it to grasp knowledge as the relationship between the information and its associated class i.e. the function that takes information as an input and returns its estimated class.
Unsupervised learning works with a set of information in which it is able to find clusters of similar items. It creates knowledge from a set of information by exposing relationships among information items and sorting them into different clusters. Reinforcement learning, finally, can be used by an agent to learn the optimal behavior to adopt in a context of interaction with an environment to maximize a user-defined reward.

Once trained, a machine learning model acts as a piece of knowledge, able to return synthetic knowledge from input information. The knowledge that is extracted through learning techniques can be further leveraged through knowledge composition methods where existing knowledge is further analyzed/collated to produce new knowledge. For instance, if a user wants to avoid traffic congestion, the system needs to first detect the congested zones and decompose the necessary factors including current location, destination, the estimated route/arrival time based on the current traffic. In this case, in addition to knowledge creation, the knowledge composition also collates some information and/or other knowledge, such as closed roads and/or construction zones, so as not to exacerbate the congestion.

Thus, the word \emph{knowledge} can be used to refer to both: (i) 
the trained model or algorithm able to synthesize sets of information into a piece of knowledge, and (ii) the generated piece of knowledge, that further abstracts the available input information. As a consequence, we believe it is necessary to take both aspects into account as we suggest a formalization of knowledge definition, storage and distribution in vehicular networks.
\begin{figure}[H]
    \centering
    \includegraphics[width=\columnwidth]{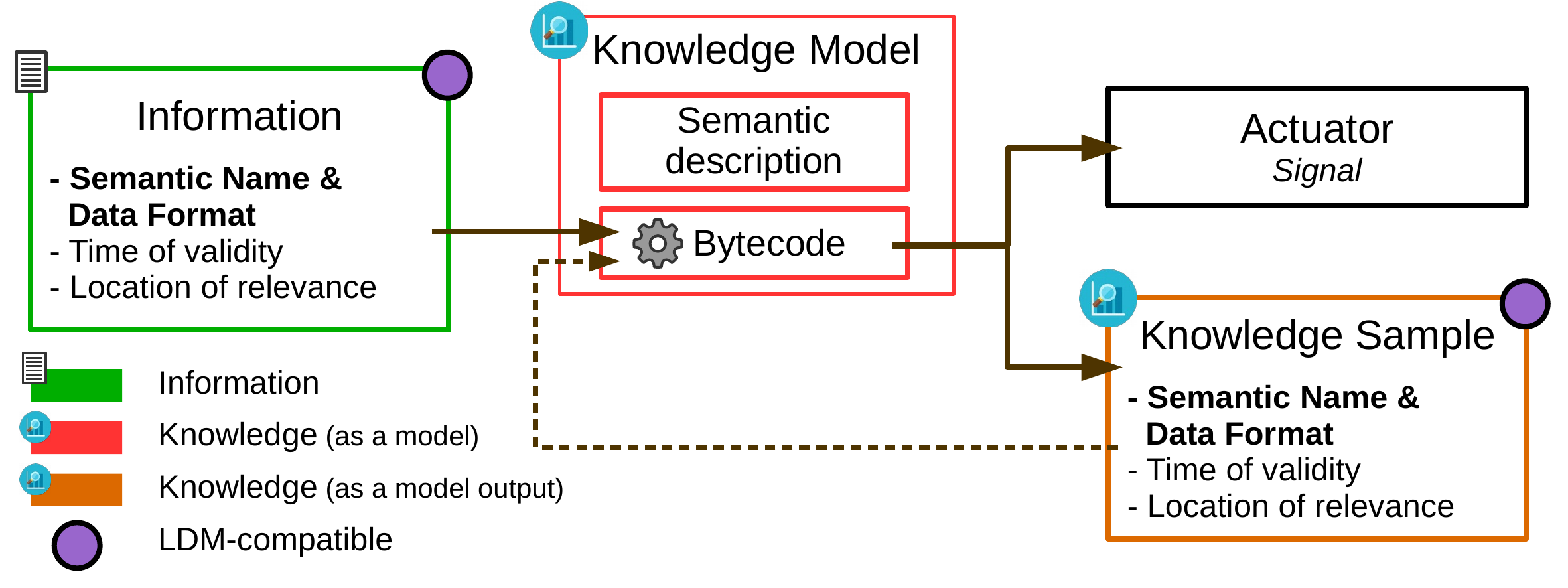}
    \caption{The Vehicular Knowledge Ecosystem}
    \label{fig:archi_vehic_knowledge}
\end{figure}
Figure~\ref{fig:archi_vehic_knowledge} shows a general picture of when and how knowledge is handled in vehicular networks for safety and driving-related applications. To the left of the figure, information is meant to be stored in the LDM, and has a time and area of validity. For the sake of interoperability with other vehicles, we consider it to be named and structured following well-known constraints, e.g. following the VSS specification.

A \emph{knowledge model}, typically implemented as a trained machine learning algorithm, takes as input several pieces of information and produces a certain piece of knowledge as an output. In Figure~\ref{fig:archi_vehic_knowledge}, we distinguish two aspects of a knowledge model: a semantic description and a bytecode. The semantic description is used to describe the necessary input, the produced output and the potential preconditions necessary to the application of the model. We define the \emph{bytecode} of a model as the executable file that produces an output from a well-formed set of input information. By executing the bytecode of a model, one can produce a \emph{knowledge sample} that, just like the information used to produce it, has a time and area of validity. Another pertinent possible output for vehicles is an actuation signal, that may be triggered in reaction to the analysed input information.

In practice, the generated \emph{knowledge sample} is structured similarly to information, the difference being that it is abstract and obtained through analysis. As a consequence, it can be fed as input to another knowledge model, generating new composed knowledge.
\begin{figure*}[t!]
    \centering
    \includegraphics[width=0.75\textwidth]{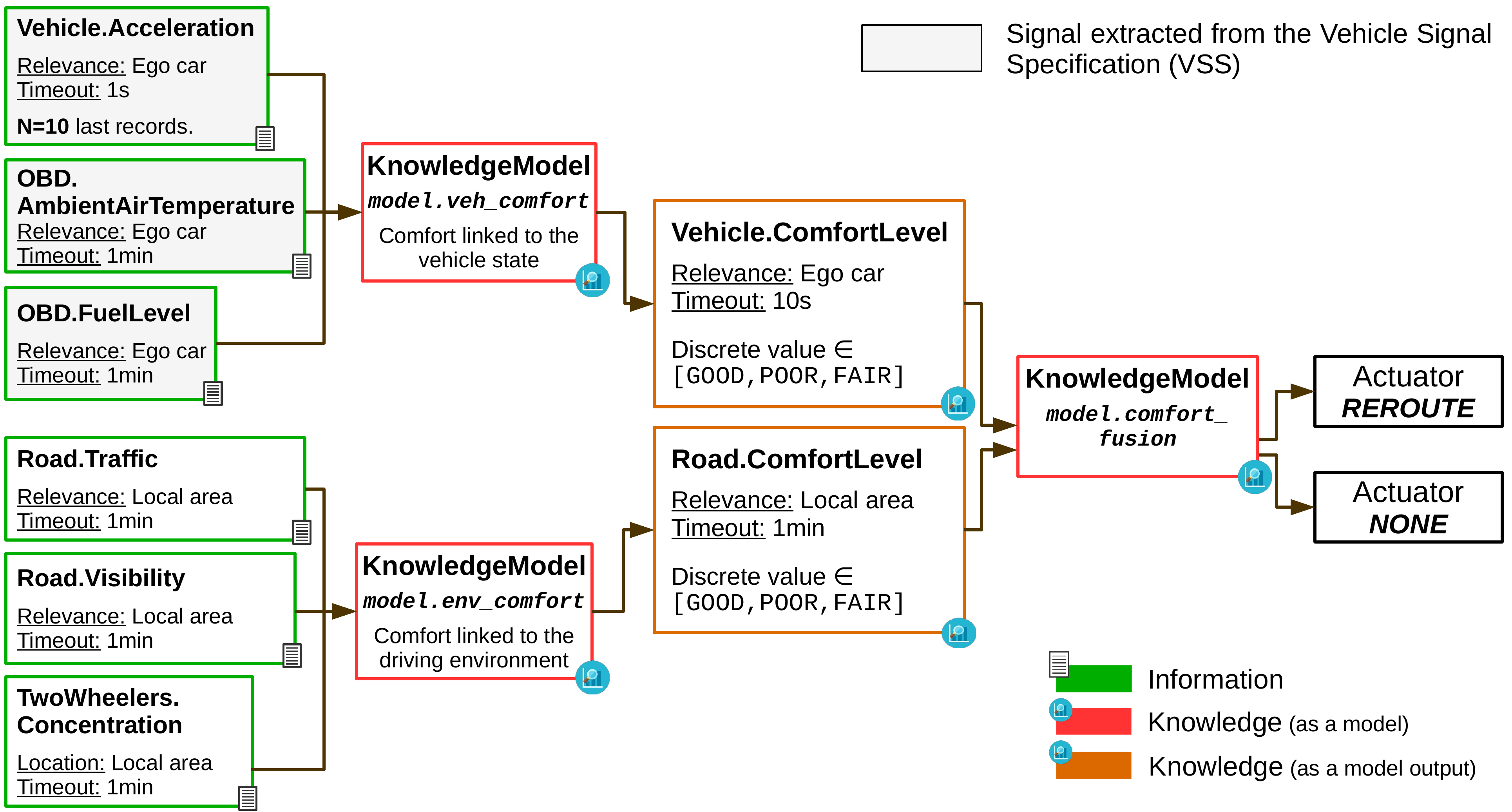}
    \caption{Knowledge Models for Passenger Comfort Estimation.}
    \label{fig:archi_applied_comfort}
\end{figure*}

Figure~\ref{fig:archi_applied_comfort} shows an application of this definition of knowledge. It is related to the estimation of passenger comfort on-board a connected vehicle. We will come back to the \texttt{model.env\_comfort} model in Section~\ref{sec:application}, as we describe an application of VKN.

\section{Aspects of Vehicular Knowledge Networking}
\label{sec:vkn}

\subsection{Knowledge Description \& Storage}

The creation of knowledge in vehicular networks takes place at two levels:
\begin{itemize}
    \item Using machine learning algorithms, automakers and organizations train and provide models capable of generating a piece of abstract knowledge from a set of input (information or \emph{knowledge samples}).
    \item The models are applied to real input, generating new \emph{knowledge samples}.
\end{itemize}

Knowledge models themselves are made of two separable components: (i) A description of the input and output parameters of the model, and (ii) the actual bytecode of the model, a program that will generate output \emph{knowledge samples} given well-formed input. Traditionally in vehicular networks, the two components are not separated. All aspects of the creation of a knowledge model, including training information gathering, model definition and training, are typically planned and organized by a single entity. The obtained model is then downloaded exclusively to the vehicle fleet of the training entity so that \emph{knowledge samples} can be locally generated.

In some cases, keeping models proprietary can be an intentional choice by automakers in order to protect their competitiveness in the market. However, in other cases, a lack of cooperation in knowledge model building may lead to an inefficient use of resources. On the one hand, similar knowledge models are likely to be independently trained by competing entities, leading to redundant computations. On the other hand, no common format to describe the input, output, and preconditions of a model are provided by training entities. This prevents the cooperative use of knowledge models by a larger number of nodes.

\begin{figure}
    \centering
    \includegraphics[width=0.8\columnwidth]{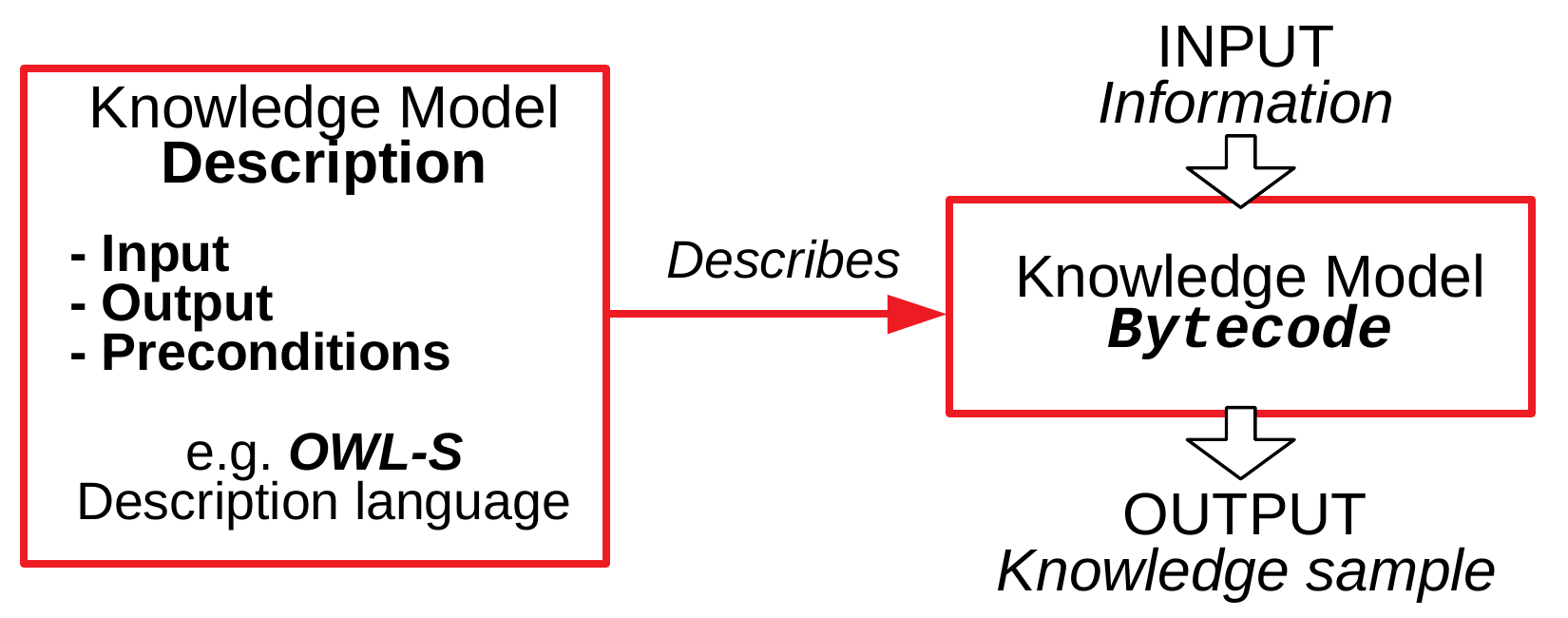}
    \caption{Separation of Model Description and Bytecode}
    \label{fig:knowledge_separation}
\end{figure}

To tackle these issues, the semantic description of a model may be physically separated from its actual execution place, as shown in Figure~\ref{fig:knowledge_separation}. The knowledge model description formally states the input, output and preconditions to the application of a given model. It is a lightweight content to be shared with multiple nodes. The knowledge model's bytecode is the implemented program whose input and output match those of the semantic description. It performs the knowledge samples creation. Even if a vehicle is not in possession of a model's bytecode, it may request knowledge creation from another node using the constraints detailed in the model description.

As a requirement for model description, the inputs and outputs of a model should be named according to standard semantics specifications such as VSS~\cite{vsso2018}. By consulting the specification for the name associated with each input or output, nodes are able to deduce the format of information and \emph{knowledge samples} required to apply the model. Moreover, preconditions to the model application may be set, e.g. conditions on the input information's age.

A possible candidate model description language matching these requirements is OWL-S, as described by Martin~\emph{et al.}~\cite{owls}. It was originally developed for automatic Web Services discovery, composition and invocation. The \emph{process model} standard of OWL-S provides a means of description for the set of input, output and preconditions of a model. Figure~\ref{fig:owls_example} provides an example of a OWL-S description of the \texttt{model.env\_comfort} model introduced in the bottom left corner of Figure~\ref{fig:archi_applied_comfort}.

\begin{figure}
\centering
\hspace*{0.05in}
\begin{tabular}{c}
\begin{lstlisting}[language=Xml]
<AtomicProcess ID="model.env_comfort">
  <hasInput resource="#traffic" />
  <hasInput resource="#visibility" />
  <hasInput resource="#twoWheelers" />
</AtomicProcess>

<Input ID="traffic">
  <parameterType resource="#Road.Traffic" />
</Input>
<Input ID="visibility">
  <parameterType resource="#Road.Visibility" />
</Input>
<Input ID="twoWheelers">
  <parameterType resource="#TwoWheelers.Concentration" />
</Input>
<Output ID="comfort">
  <parameterType resource="#Road.ComfortLevel" />
</Output>
\end{lstlisting}
\end{tabular}
\caption{Comfort Model Description Structure in OWL-S}
\label{fig:owls_example}
\end{figure}

\begin{figure}
    \centering
    \includegraphics[width=\columnwidth]{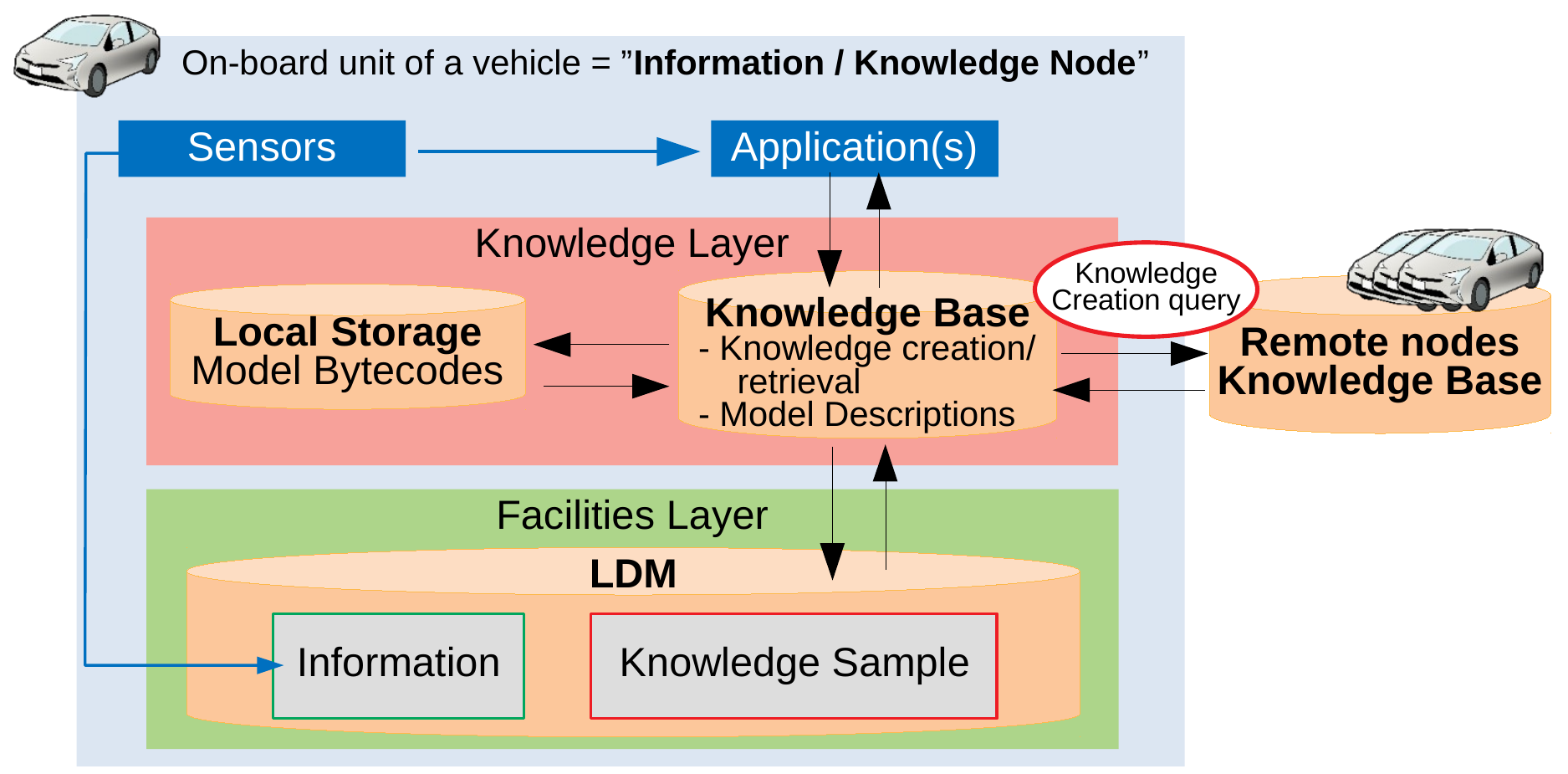}
    \caption{On-Board Storage of Knowledge and Information}
    \label{fig:knowledge_storage}
\end{figure}

As part of VKN, we suggest separating the storage of models' descriptions and bytecodes. On-board each connected vehicle, as shown in Figure~\ref{fig:knowledge_storage}, we suggest adding a \emph{knowledge} layer on top of the ITS facilities layer containing the LDM:
\begin{itemize}
    \item In a Knowledge Base (KB), a list of known knowledge model descriptions are stored.
    \item A local storage in the knowledge layer may store knowledge model bytecodes. It is responsible for the actual execution of the knowledge creation process. The stored bytecodes are independent from the model descriptions stored in the KB.
    \item The generated \emph{knowledge samples} are stored in the LDM along with traditional information.
\end{itemize}

The KB is responsible for knowledge creation, and knowledge retrieval when the knowledge is remotely created. When creating knowledge, the KB looks up in the LDM for the necessary input items. Then, if the required model bytecode is stored locally, it is executed immediately. The KB also communicates with other nodes' KBs to reach for non locally-stored bytecodes. This allows for a flexible framework, allowing to expand or limit the creation and distribution of knowledge within a certain group of vehicles as needed. The modalities for such remote knowledge creation will be further developed in the knowledge distribution section.

\subsection{Knowledge Distribution}

\begin{figure}
    \centering
    \includegraphics[width=0.9\columnwidth]{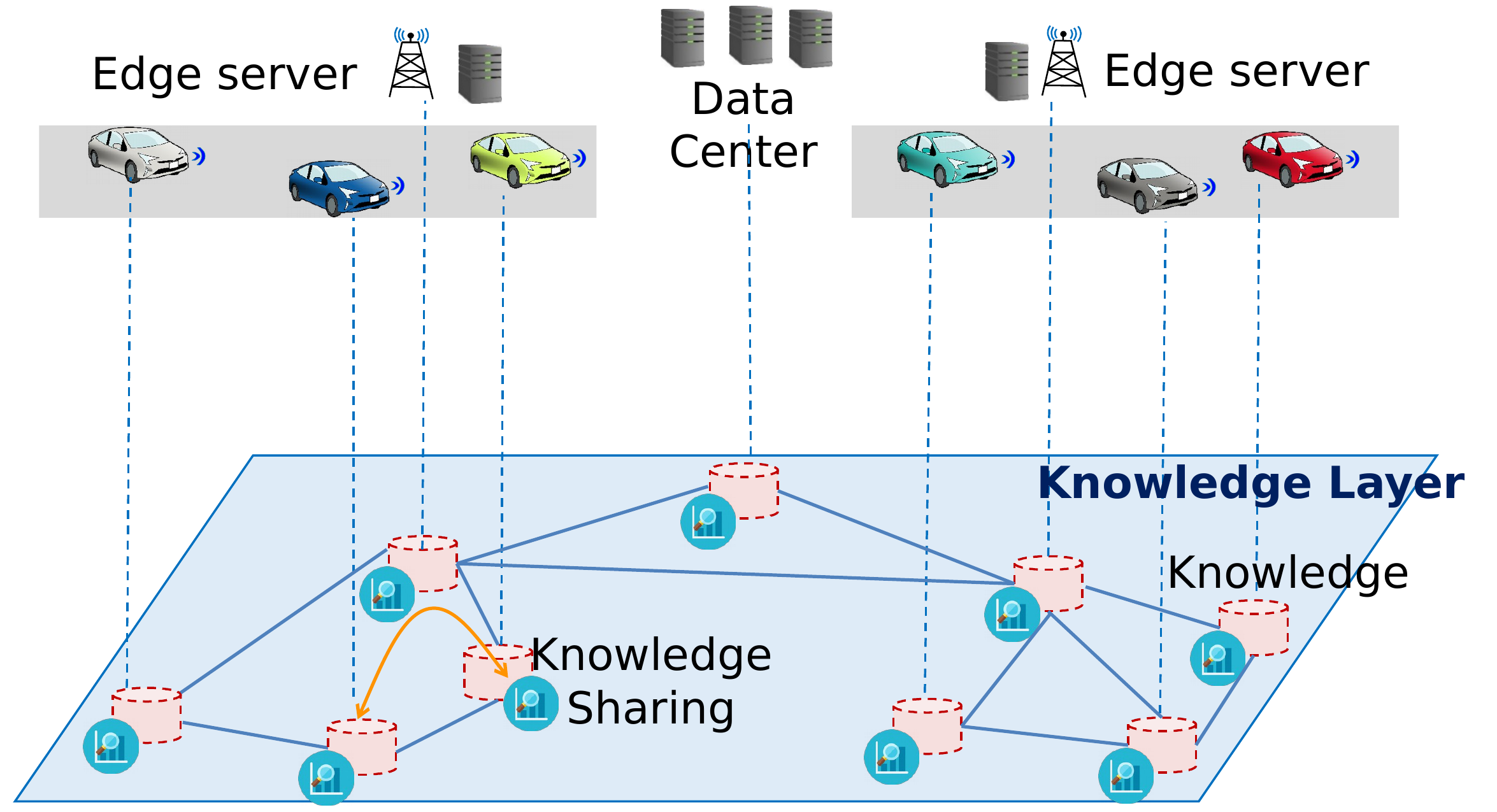}
    \caption{Architecture of Vehicular Knowledge Networking}
    \label{fig:knowledge_layer_networking}
\end{figure}

As we separate knowledge models' bytecodes and their description, a structural need appears for the distribution of knowledge. As shown in Figure~\ref{fig:knowledge_layer_networking}, nodes of the vehicular network are inter-connected and knowledge creation may be the product of a cooperation between multiple nodes.

The input information or \emph{knowledge samples} required for knowledge creation through a knowledge model are spread across the vehicular network. As a consequence, centralizing all the input content required for knowledge creation in a single node can be inefficient. VKN makes knowledge models mutually understandable. It brings the opportunity not only to share models themselves, but also to outsource knowledge creation to remote nodes fitted with pertinent input. By cooperatively creating knowledge among well-chosen nodes, the amount of potentially heavyweight model inputs transmitted within the vehicular network can be reduced, saving time and load.

In order to achieve knowledge distribution, a \emph{Vehicular Knowledge Querying Language} (VKQL) must be defined to:
\begin{enumerate}
    \item Lookup and retrieve the available knowledge models descriptions, their required input, output and preconditions.
    \item Spread knowledge model bytecodes efficiently.
    \item Delegate knowledge creation to remote nodes.
\end{enumerate}
An example of the delegation of knowledge creation to remote nodes and its potential benefits is developed in Section~\ref{sec:application}, through the example of passenger comfort-based rerouting.

Finally, a balance must be found between widespread and too scarce distribution of bytecodes. The former would increase load in the network and negate the point of separating descriptions and bytecodes, while the latter would make access to bytecodes too difficult, increasing delay. Mechanisms to cache knowledge close to the vehicles could make use of technologies such as edge computing units~\cite{mach2017} or vehicular micro-clouds~\cite{VC5}.

\section{Application: Passenger Comfort-based rerouting}
\label{sec:application}

As an application of VKN and to show potential benefits for vehicular networks, we investigate on the use case of \emph{passenger comfort-based rerouting}.
The scenario is the following: a Highly-Automated Vehicle (HAV) $V_{ego}$ seeks to provide the most \emph{comfortable} driving itinerary to its passengers.

\subsection{Comfort Knowledge Model}

Under certain driving situations, as surveyed by Xin~\emph{et al.}~\cite{xin2019}, a HAV's autonomous driving ability can be challenged. In somes cases, it can lead to a take over by a human driver, which can be regarded as a factor of discomfort for passengers.

Based on those factors and as an example, we defined a simplified knowledge model \texttt{model.env\_comfort} to determine the level of \emph{passenger comfort} in a given area, as introduced in the bottom left corner of Figure~\ref{fig:archi_applied_comfort}. The model reads a set of area-related input with semantically-defined names:
\begin{enumerate}
    \item The current traffic conditions,\\$tr \in $ \texttt{Road.Traffic} $\equiv$ [\texttt{FLUID}, \texttt{CONGESTED}].
    
    \item The visibility in the area, $v \in $ \texttt{Road.Visibility} $\equiv$ [\texttt{CLEAR}, \texttt{OBSTRUCTED}].
    
    \item The concentration of two-wheelers in the surroundings, $c_{tw} \in $ \texttt{TwoWheelers.Concentration} $\equiv$ [\texttt{HIGH}, \texttt{MEDIUM}, \texttt{LOW}].
\end{enumerate}

Based on this input, the model returns a discrete qualification of the level of comfort associated with driving in the area, as $cft \in $ \texttt{Road.ComfortLevel} $\equiv$ [\texttt{GOOD}, \texttt{FAIR}, \texttt{POOR}].
To demonstrate what a simplified version of a knowledge bytecode would be, we provide a simple pseudocode implementation of the model in Algorithm~\ref{algo:the_algo}.

\begin{algorithm}[H]
\small
\caption{Simplified algorithm to compute comfort from environmental parameters}
\label{algo:the_algo}

 \hspace*{\algorithmicindent} \textbf{Input} \\
 \hspace*{\algorithmicindent}  $c_{tw} \in $ [\texttt{HIGH}, \texttt{MEDIUM}, \texttt{LOW}] \\
 \hspace*{\algorithmicindent} $v \in $ [\texttt{CLEAR}, \texttt{OBSTRUCTED}] \\
 \hspace*{\algorithmicindent} $tr \in $ [\texttt{FLUID}, \texttt{CONGESTED}] \\
 \hspace*{\algorithmicindent} \textbf{Output} \\
 \hspace*{\algorithmicindent} $cft \in $ [\texttt{GOOD}, \texttt{FAIR}, \texttt{POOR}]
\begin{algorithmic}[1]

\IF{$c_{tw}$ = \texttt{LOW} \AND $v =$ \texttt{CLEAR} \AND $tr$ = \texttt{FLUID}}
\STATE $cft \leftarrow$ \texttt{GOOD}
\ELSIF{$c_{tw}$ = \texttt{HIGH}}
\STATE $cft \leftarrow$ \texttt{POOR}
\ELSE
\STATE $cft \leftarrow$ \texttt{FAIR}
\ENDIF
\end{algorithmic}
\end{algorithm}

\subsection{Comfort Retrieval Scenario}

\begin{figure}
    \centering
    \includegraphics[width=\columnwidth]{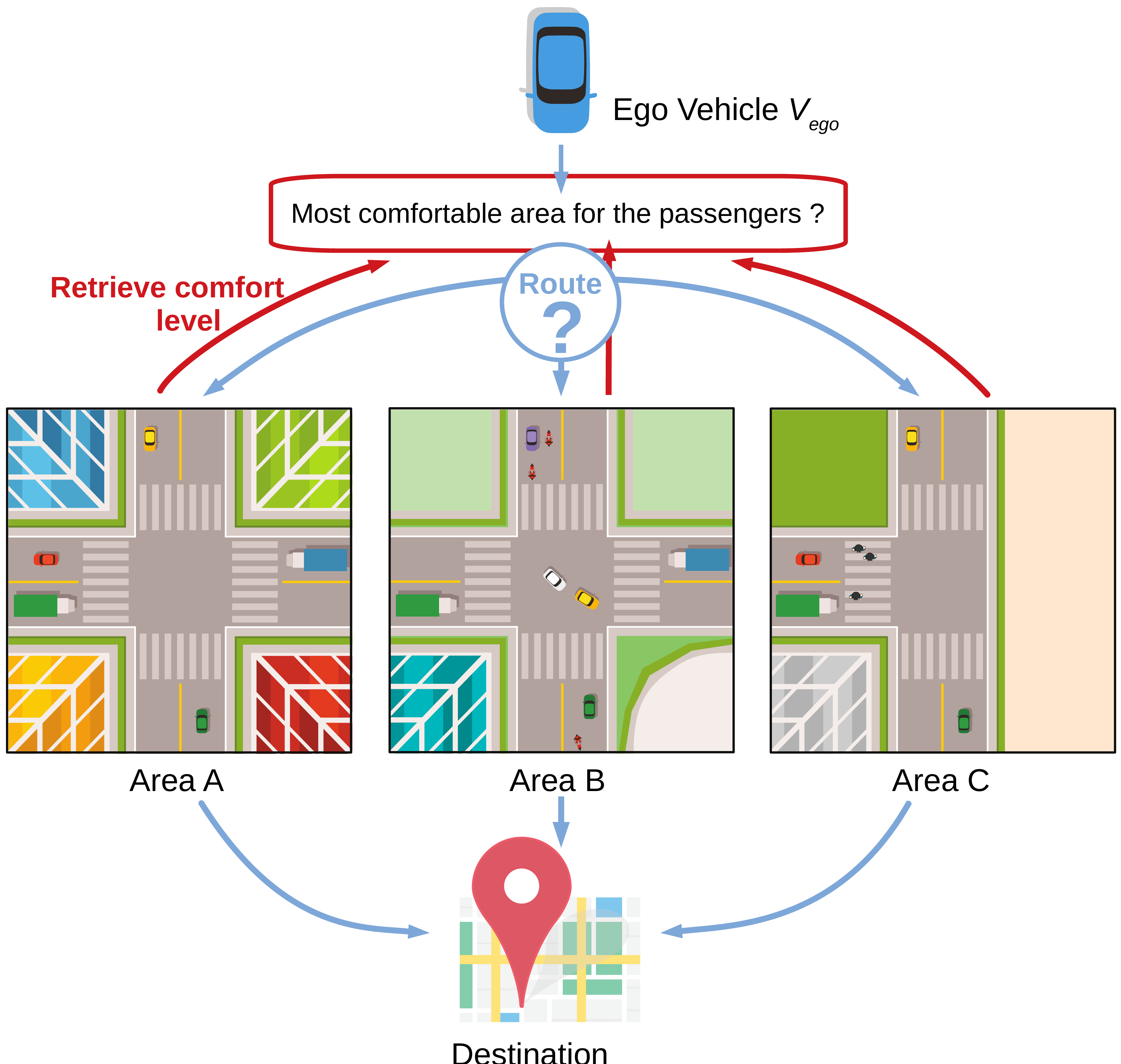}
    \caption{Illustration of the Comfort-based Rerouting Problem.}
    \label{fig:glob_scenario}
\end{figure}

The vehicle $V_{ego}$ has three itinerary options to reach its destination: crossing remote area $A$, $B$, or $C$. In order to compute the most comfortable itinerary for its passengers and take a decision on which route to take, $V_{ego}$ wishes to obtain a \texttt{Road.ComfortLevel} knowledge sample for each of these areas. Figure~\ref{fig:glob_scenario} illustrates the problem of the scenario.

Once $V_{ego}$ has obtained the comfort level on each of the areas $A$, $B$, and $C$, it is able to take an educated decision on what itinerary to follow. As an illustration, we describe the process of retrieving the driving comfort level from Area $A$. The problem is as follows: $V_{ego}$ is currently driving $N \geq 5km$ away from Area $A$, and wishes to obtain the driving comfort level in $A$ prior to crossing it, in order to potentially reroute beforehand.

In a traditional, information-centric approach lacking standards for knowledge exchange, a typical solution to this problem would be as follows:
\begin{enumerate}
  \setcounter{enumi}{-1}
    \item The comfort knowledge model is downloaded to $V_{ego}$.
    \item $V_{ego}$ transmits an information request via ICN to the target area.
    \item The information acquisition is sensed by a node in the target area and retrieved to $V_{ego}$.
    \item The remote comfort level is computed on-board $V_{ego}$.
\end{enumerate}
The input information is typically heavier than the produced knowledge. As a consequence, this approach increases load and delay in vehicular networks compared with a direct transfer of the knowledge.

\begin{figure}
    \centering
    \includegraphics[width=\columnwidth]{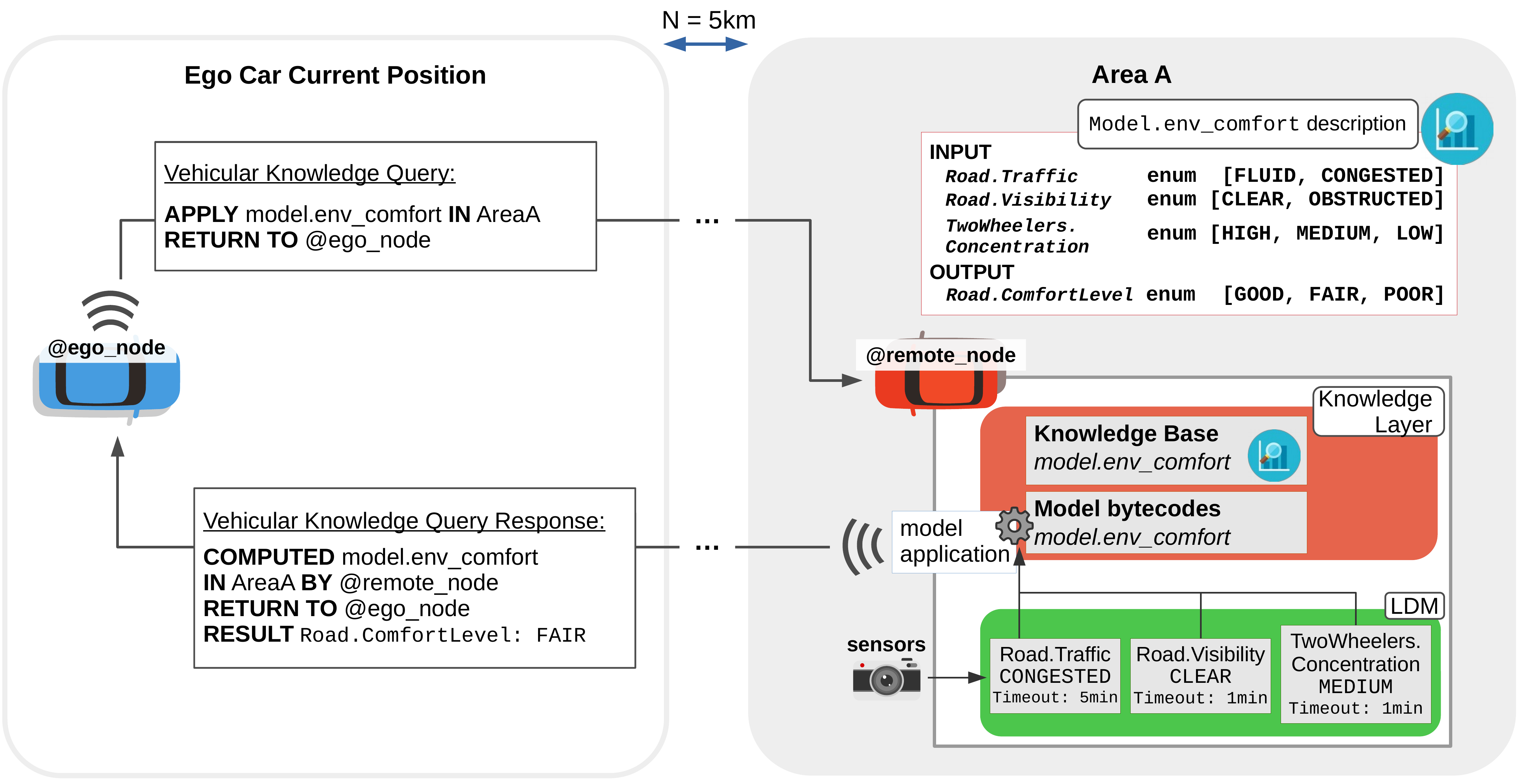}
    \caption{Summary of the Comfort Level Retrieval in a Remote Area using VKN}
    \label{fig:scenario}
\end{figure}

In a novel solution, making use of VKN, we suggest to outsource the production of knowledge to a node in the target area. As shown in Figure~\ref{fig:scenario}, instead of transmitting a request for input information to nodes in the target area, $V_{ego}$ directly transmits a request for the comfort level knowledge in the area. The knowledge creation request is then forwarded to a node in the target area in possession of the required \texttt{model.env\_comfort} model's bytecode. In turn, the remote node locally computes the comfort knowledge using input information stored in its LDM and obtained through sensors. After the knowledge has been produced, it is directly returned to $V_{ego}$. Knowledge is typically much lighter than the input it was extracted from. In turn, this allows to save load and delay in vehicular networks.

\section{Research Applicability}
\label{sec:opportunities}

We described an application of cooperative \emph{knowledge samples} creation. By remotely running a knowledge model's bytecode in the area where its input information is sourced, unnecessary transfers of information are avoided to the benefit of knowledge.

Similarly, mechanisms have been defined in the literature to train \emph{knowledge models} themselves while avoiding the transmission of training information for privacy and efficiency concerns. Federated Machine Learning (FML) is an open research topic in which multiple nodes cooperatively train and update a shared model without exchanging actual training information~\cite{fml2019}. A centralized service provider selects a set of client nodes to cooperatively train a model. A current state of the model to be trained is downloaded to each client. Then, each client locally performs a few training steps of the model with local information before sending the updated model state back to the server. The server then aggregates the updated client model states to compute a global update of the model, taking advantage of the local training of each client.

An issue of this approach is that it is not trivial to ensure that all local nodes are interested in training and using the same model. Before being able to start the training, local nodes should be able to determine who among their neighbors is in possession of what type of model and has access to what kind of information.
We believe this to be a pertinent use case for VKN, as a protocol of discovery of the knowledge already possessed by one's neighbor is needed to initialize a session of FML.

\section{Conclusion}
\label{sec:conclusion}

Vehicular networks have been extensively studied in the past years. Several standards have been developed to store and share information. However, challenges remain to transition from an information-centric networking model to a model where common standards for knowledge characterization, description, storage and sharing allow nodes in vehicular networks to take full advantage of data-driven AI techniques. In this paper, using a common definition of knowledge, we determined under what forms it exists in vehicular networks, allowing us to concretely propose a structure for knowledge description, storage and sharing. Through a passenger comfort-based rerouting application, we exemplified the concept and showed potential performance improvements. Finally, we note the potential benefits of Vehicular Knowledge Networking for the open topic of Federated Machine Learning. Future work will focus on implementing, simulating and measuring the benefits of using VKN, with regards to traditional information-centric models.

\balance
\printbibliography
\end{document}